\documentclass[natbib209]{emulateapj}

\usepackage{amsmath}
\usepackage{epsfig}

\newcommand{\pd}{\partial}

\newcommand{\del}{\mbox{\boldmath $\nabla$}}
\newcommand{\curl}{\mbox{\boldmath $\nabla \times$}}

\newcommand{\cross}{\mbox{\boldmath $\times$}}

\newcommand{\vv}{{\bf v}}
\newcommand{\BB}{{\bf B}}

\newcommand{\rh}{\overline{\rho}}

\newcommand{\uvp}{\mbox{\boldmath $\hat{\phi}$}}

\shorttitle{Convective Babcock-Leighton Dynamo Models}
\shortauthors{Miesch \& Brown}

\begin{document}

\title{Convective Babcock-Leighton Dynamo Models}

\author{Mark S. Miesch$^1$ and Benjamin P. Brown$^2$}
\affil{$^1$High Altitude Observatory, National Center for Atmospheric Research, Boulder, CO, 80307-3000, USA: miesch@ucar.edu}
\affil{$^2$Department of Astronomy and Center for Magnetic Self-Organization in Laboratory 
and Astrophysical Plasmas, University of Wisconsin, 1150 University Avenue, Madison, WI 53706, USA}

\begin{abstract}
We present the first global, three-dimensional simulations of solar/stellar
convection that take into account the influence of magnetic flux emergence
by means of the Babcock-Leighton (BL) mechanism.  We have shown that the 
inclusion of a BL poloidal source term in a convection simulation can 
promote cyclic activity in an otherwise steady dynamo.  Some cycle properties 
are reminiscent of solar observations, such as the equatorward propagation of 
toroidal flux near the base of the convection zone.  However, the cycle period 
in this young sun (rotating three times faster than the solar rate) is very 
short ($\sim$ 6 months) and it is unclear whether much longer cycles
may be achieved within this modeling framework, given the high efficiency
of field generation and transport by the convection. 
Even so, the incorporation of mean-field parameterizations in 
3D convection simulations to account for elusive processes such 
as flux emergence may well prove useful in the future modeling of
solar and stellar activity cycles.
\end{abstract}


\section{Introduction}\label{sec:intro}

The emergence of magnetic flux through the solar photosphere regulates solar
variability and powers space weather.  It is clear that this flux originates
in the solar interior and is produced by the solar dynamo.  However, it is 
currently unclear what role flux emergence plays in establishing the 22-year 
solar activity cycle.  Is it an essential ingredient or merely a superficial 
by-product of deeper-seated dynamics?

One of the principle means by which flux emergence may act an an
essential ingredient in the operation of the solar dynamo is through
the so-called Babcock-Leighton (BL) mechanism \citep{babco61,leigh64}.
The BL mechanism arises through the dynamics of flux emergence.  As a
buoyant flux tube rises through the convection zone, the Coriolis
force induces a twist in the axis of the tube that is manifested upon
emergence as a poleward displacement between the trailing and leading
edges of a bipolar active region.  When the active region subsequently
fragments and disperses due to surface convection and meridional flow,
the redistribution of vertical flux induces a net electromotive force
(emf) that converts mean toroidal field to poloidal field.  Although 
doubts remain about the viability of the BL mechanism as the principle 
source of poloidal flux, its empirical foundation and robustness have 
made it an integral element in many recent mean-field dynamo models 
of the solar cycle \cite[reviewed by][]{dikpa09,charb10}.

\begin{figure*}
\centerline{\epsfig{file=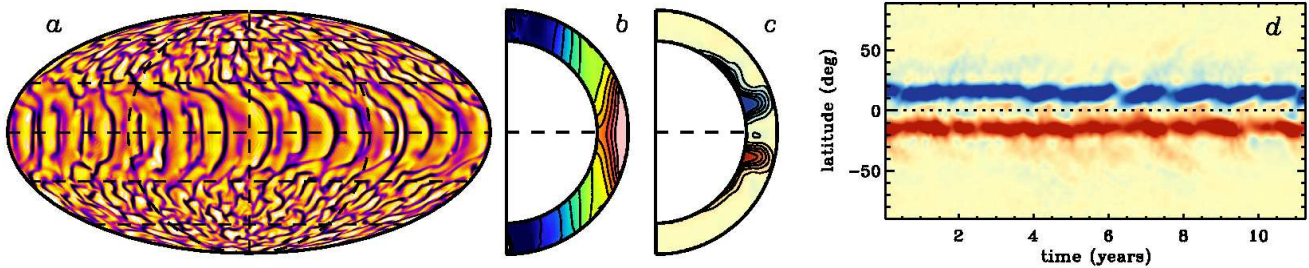,width=\linewidth}}
\caption{Progenitor simulation, Case D3, from \cite{brown10}.
($a$) Radial velocity in a Mollweide projection near the
top of the computational domain ($r = 0.953R$, $t = 2$ yr, color table
saturation $\pm$ 70 m s$^{-1}$, yellow/orange upflow, blue/black downflow).
($b$) Angular velocity $\Omega/2\pi$, averaged over 1200 days 
(7.8-11.3 yrs, saturation 1140--1320 nHz, white/red fast, blue/black slow).
($c$) Mean toroidal magnetic field $\left<B_\phi\right>$ averaged over 
longitude and time (15 day average near $t = 2$ yrs, saturation $\pm$ 8 kG, 
red positive, blue negative).  ($d$) $\left<B_\phi\right>$ in the mid
convection zone ($r = 0.84R$) versus latitude and time (colors as in $c$, 
saturation $\pm$ 4 kG).
\label{fig:D3}}
\end{figure*}

Current simulations of global solar and stellar convection do exhibit sustained 
dynamo action and, depending on the parameter regime, can produce magnetic
cycles \citep{ghiza10,racin11,brown11}.  Yet, these simulations do not have sufficient 
resolution to reliably capture flux emergence or the BL mechanism. \cite{nelso11} 
recently reported the first convective dynamo simulation to exhibit the spontaneous,
self-consistent generation of buoyant magnetic flux structures generated by 
convectively-driven rotational shear.  However, even these cannot realistically simulate
the subtle, multi-scale dynamics of flux emergence and dispersal that underlies the 
BL mechanism.  This would require (1) very low diffusion to form concentrated flux 
structures, (2) very high resolution to capture the destabilization and coherent 
rise of those structures, and (3) a realistic depiction of surface convection, 
meridional flow, and radiative transfer in order to capture the emergence,
coalescence, fragmentation, and dispersal of bipolar active regions \citep{cheun10}.  
Each of these is a formidable computational challenge in its own right, stetching 
the limits of modern supercomputers.  No global model is capable of unifying 
convective dynamos and the BL mechanism through direct numerical simulation.

Thus, the presence of magnetic cycles in global convection simulations
demonstrates that the BL mechanism is not a necessary ingredient for
cyclic activity.  How this may or may not apply to the solar dynamo is
a complex, unresolved issue and we make no attempt at a comprehensive
discussion here.  Our purpose is rather to investigate how flux emergence
may alter the behavior of a convective dynamo by means of the BL mechanism.
The BL mechanism is modeled using a mean-field parameterization
intended to mimic dynamics that cannot be explicitly captured by the
simulation itself for the reasons outlined above.  Our approach is
described in \S\ref{model} and simulation results are presented in
\S\ref{results}, along with interpretive discussion. 

Before proceeding, it is worth emphasizing up front that the simulations 
we consider here are of a solar-like star rotating three times more
rapidly than the Sun (3$\Omega_\odot$).  This is done because it is 
a tidy numerical experiment; without BL forcing, this dynamo builds 
strong large-scale fields that do not undergo cycles.  Other parameter
regimes, including those at the solar rotation rate, will be considered
in future work.

\section{Model}\label{model}

The starting point for our investigation is the convective dynamo simulation that 
we refer to as case D3 and that is described in detail by \cite{brown10}.
We refer
the reader to that paper for further information on the set up and results of the 
simulation as well as the ASH (Anelastic Spherical Harmonic) code that is used to 
solve the equations of magnetohydrodynamics (MHD) in a rotating spherical shell
under the anelastic approximation.  Solar values are used for the luminosity
and the background stratification but the rotation rate is a factor of three
faster than the Sun, as mentioned above.  The spatial resolution of all simulations 
reported in this paper is 96 $\times$ 256 $\times$ 512 ($r$, $\theta$, $\phi$).  
The simulation domain spans from $r_1 =  0.718 R$ to $r_2 = 0.966 R$, where $R$ is 
the solar radius.

The salient feature of Case D3 that we are interested in here is the
presence of persistent toroidal field structures that we term magnetic
wreaths.  The convection exhibits an intricate, evolving small-scale
structure (Fig.\ \ref{fig:D3}$a$) but produces a substantial
differential rotation (Fig.\ \ref{fig:D3}$b$) that in turn promotes
the generation of prominent magnetic wreaths (Fig.\ \ref{fig:D3}$c$).
The mean toroidal field is approximately symmetric about the equator,
with one wreath in each hemisphere of opposite polarity.  Although
continual buffeting by convective motions induces non-axisymmetric and
temporal fluctuations, the mean field is remarkably persistent,
retaining its essential structure indefinitely (Fig.\ \ref{fig:D3}$d$).  
After they are established, the wreaths persist for at least 60 years 
(the duration of the simulation) with no sign of abating \citep{brown10}.  
This time interval is much longer than the rotation period (9.3 days), 
the convective turnover time scale (about 20 days) and the ohmic 
diffusion time (about 3.6 years).

The simulations described in this paper are all restarted from the
same iteration of Case D3, defined as time $t = 0$.  This includes 
the unmodified D3 run shown in Figure \ref{fig:D3}, which was 
continued beyond the restart iteration for comparison purposes.  
The absence of initial transients in Fig.\ \ref{fig:D3} demonstrates 
that this simulation has reached a statistically steady state by
the mutual reference time $t = 0$.

The anelastic MHD equations for the conservation
of mass, momentum, and thermal energy are solved with no modification.
The only difference between the simulations presented here and
that presented in \cite{brown10} is in the magnetic induction
equation where we add an additional poloidal source term 
$S(r,\theta)$ as follows:
\begin{equation}\label{eq:indy}
\frac{\pd \BB}{\pd t} = \curl \left(
\vv \cross \BB - \eta \curl \BB + S \uvp\right)
\end{equation}
The additional term is intended to mimic the generation of mean
poloidal field by the BL mechanism.  Following the mean-field BL
dynamo model of \citet{rempe06}, we choose
\begin{equation}
S(r,\theta) = \alpha f(r) g(\theta) \hat{B}_\phi  
\end{equation}
where $f(r)$ and $g(\theta)$ are radial and latitudinal profiles
\begin{equation}\label{frad}
f(r) = \mbox{max}\left[ 0, 1 - \frac{(r-r_2)^2}{d^2}\right]
\end{equation}
\begin{equation}
g(\theta) = \frac{3 \sqrt{3}}{2} ~ \sin^2\theta \cos \theta
\end{equation}
and $\hat{B}_\phi$ is a measure of the mean toroidal flux at the
base of the convection zone
\begin{equation}\label{Bhat}
\hat{B}_\phi = \int_{r_1}^{r_b} h(r) \left<B_\phi\right> dr ~~~.
\end{equation}
Brackets $<>$ denote an average over longitude and $h$ is an averaging 
kernal given by $h(r) = h_0 \left(r - r_1\right)\left(r_b - r\right)$.
The integration is confined to a region near the base of the convection
zone, below $r_b = 0.79 R$ and the normalization $h_0$ is defined 
such that $\int_{r_1}^{r_b} h(r) dr = 1$.

Note that the radial profile in equation (\ref{frad}) is nonzero
only near the top of the convection zone, for $r > r_2 - d$.
We choose $d = 20$ Mm so the poloidal source operates above
$r = 0.937 R$.  Thus, the BL term is nonlocal in the sense
that the poloidal source near the surface is proportional 
to the mean toroidal flux near the base of the convection 
zone.  This is typical of BL dynamo models \citep[e.g.][]{rempe06}.

The amplitude of the BL term, $\alpha$, includes an algebraic
quenching of the form $\alpha = \alpha_0 (1 + B_t^2/B_q^2)^{-1}$
where $B_q = 1$ MG is the quenching field strength and
$B_t^2 = (1/2) \int_0^\pi \hat{B}^2 \sin\theta d\theta$.
In practice the fields generated are much less than $B_q$ so
the quenching plays little role.

The magnetic diffusivity $\eta$ in all simulations is the same as
in Case D3, varying from 1.56-7.69 $\times 10^{12}$ cm$^2$ s$^{-1}$ 
from the bottom to the top of the convection zone, $\propto 
\rh^{-1/2}$, where $\rh$ is the background density.

The objective of this paper is to vary the amplitude of the BL term
$\alpha_0$ in order to investigate how flux emergence may alter the
nature of the dynamo.  We emphasize again that the 3D MHD equations 
are unmodified apart from the $S(r,\theta)$ term in eq.\ (\ref{eq:indy})
so setting $\alpha_0 = 0$ reproduces Case D3 as shown in Figure \ref{fig:D3}
and as described at length in \citet{brown10}.

\begin{figure}
\centerline{\epsfig{file=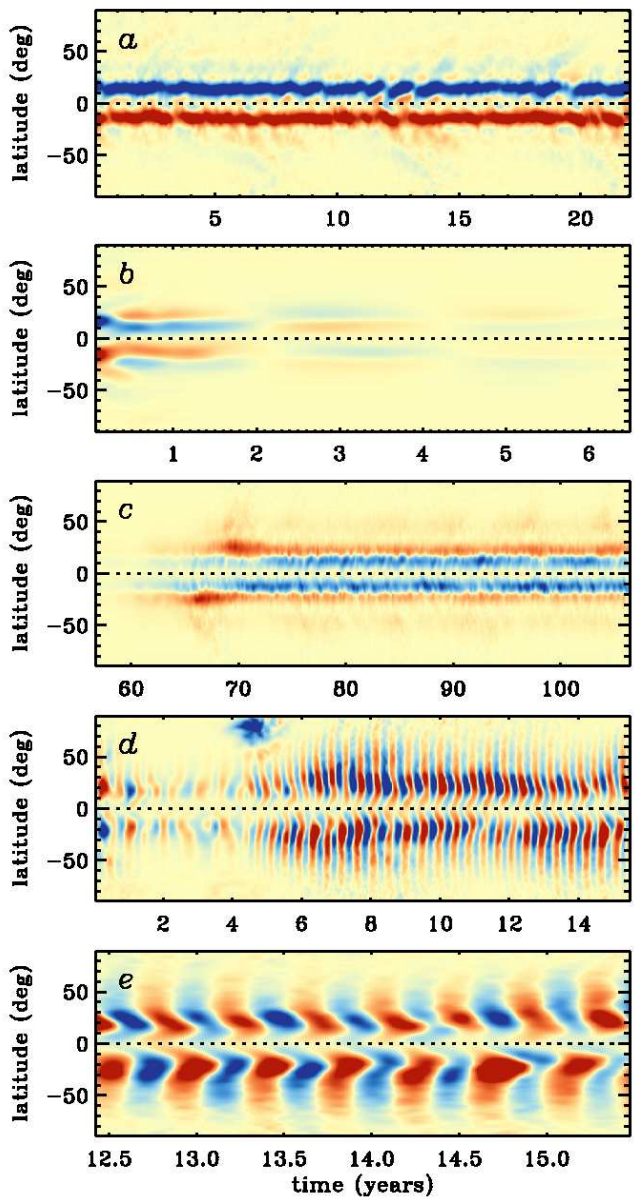,width=3.5in}}
\caption{Mean toroidal field as a function of latitude and time as 
in Fig.\ \ref{fig:D3}$d$ (same radial level, color scale, saturation
$\pm$ 4 kG) with ($a$) $\alpha_0 = $ 1 m s$^{-1}$, ($b$, $c$) 
$\alpha_0 = $ 10 m s$^{-1}$, and ($d$, $e$) $\alpha_0 = $ 100 m s$^{-1}$.  
Frame ($c$) represents a continuation of frame ($b$) at much later times.
Frame ($e$) is a zoomed-in portion of frame ($d$) highlighting
the magnetic cycles.\label{fig:bflys}}
\end{figure}

\section{Results}\label{results}

Figure \ref{fig:bflys} shows ``butterfly'' diagrams (latitude-time
plots of the mean toroidal field near the base of the convection
zone) for a series of simulations with progressively increasing 
values of the BL forcing amplitude $\alpha_0$.  For 
$\alpha_0 = $ 1 m s$^{-1}$ the BL term is too weak to sigificantly
influence the operation of the dynamo (Fig.\ \ref{fig:bflys}$a$).
The axisymmteric component of the poloidal field near the surface
does increase but not enough to effect the maintenance of persistent
wreaths in the lower convection zone.

For $\alpha_0 = $ 10 m s$^{-1}$ the results are very different.
The wreaths essentially vanish within a few years of simulation 
time (Fig.\ \ref{fig:bflys}$b$).  This can be attributed to the
differential rotation operating on the mean poloidal field generated
by the BL mechanism via the $\Omega$-effect.  The sense of the
poloidal field produced by the BL term is such that the $\Omega$-effect
generates toroidal flux in the upper convection zone that is of 
opposite polarity to the sense of the wreaths.  Convection 
rapidly mixes this flux, bringing together opposite polarities
that are annihilated through ohmic dissipation.  The magnetic 
energy in the mean fields drops rapidly, decaying by a 
factor of 10$^6$ by $t \sim $ 30 yrs.

Yet, beyond about 30 years, the magnetic energy begins to 
rise again and the dynamo is reborn, saturating by about 
$t \sim $ 75 yrs (Fig.\ \ref{fig:bflys}$c$) with a very 
different structure.  Persistent toroidal wreaths are
again present but they are symmetric about the equator,
with two wreaths per hemisphere and a somewhat lower
magnetic energy (about 27\% less than in Case D3).

Increasing $\alpha_0$ by another order of magnitude produces 
prominent magnetic cycles (Fig.\ \ref{fig:bflys}$d$).
Toroidal wreaths form at mid latitudes in each hemisphere
and propagate toward the equator, reminiscent of the solar
butterfy diagram (Fig.\ \ref{fig:bflys}$e$).  However, the
cycle period is much shorter than in the Sun; about 6 months
compared to 22 years.  Nonlinear modulation of the cycle
is evident, with waxing and waning amplitudes and transient,
weaker substructure in the butterfly diagram near the equator.

It is clear from Fig.\ \ref{fig:bflys}$e$ that reversals in 
the two hemispheres are not precisely sychronized.  A more 
careful analysis reveals that the phase difference shifts
over time, suggesting that the two hemispheres are to some
extent decoupled.  For example, the northern hemisphere leads
the southern hemisphere from $t \sim $ 4.5 -- 11.5 yrs
but the reverse is true for $t \sim $ 12 -- 14.5 yrs.
Similar shifts in the hemispheric phase difference (albeit 
less pronounced) have been reported in sunspot records;
in particular, the southern hemisphere was apparently 
leading in cycles 18-20 while the northern was leading 
in cycles 21-23 \citep{mcint12}.

\begin{figure}
\centerline{\epsfig{file=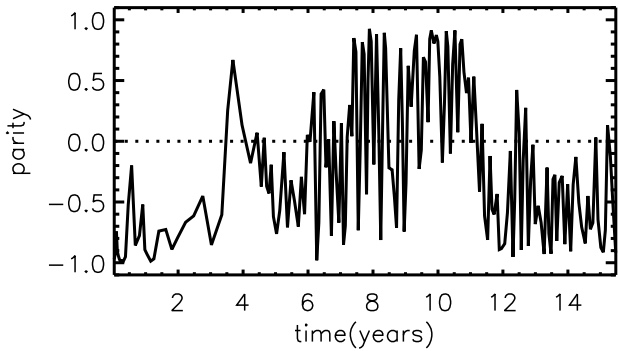,width=3.5in}}
\caption{Parity of the mean toroidal field as a function of time
for the simulation shown in Fig.\ \ref{fig:bflys}$d$,$e$ 
($\alpha_0 = $ 100 m s$^{-1}$).  \label{fig:parity}}
\end{figure}

The symmetry of the dynamo about the equator can be quantified
by the parity ${\cal P} = (B_s^2 - B_a^2)/(B_s^2 + B_a^2)$, where
$B_s$ and $B_a$ are the symmetric and antisymmtric components
of $\left<B_\phi\right>$ respectively (sampled at $r = 0.84 R$).
The parity of Case D3 and for $\alpha_0 = $ 1 m s$^{-1}$ ranges 
between -0.5 and -1 while that for $\alpha_0$ = 10 m s$^{-1}$ 
ranges between positive 0.5-1.  By contrast, the cyclic 
dynamo ($\alpha_0 = $ 100 m s$^{-1}$) shifts between positive 
and negative parity as time proceeds.  It does not exhibit the 
high synchronization of the solar cycle which exhibits a persistent 
negative parity.

As expected, the addition of a BL term has a substantial influence on
the amplitude and structure of the mean poloidal field.  Without it,
the poloidal field has a roughly octupolar structure, such that
$\left<B_r\right>$ is radially outward near the north pole, inward near
the core of the northern wreath, and antisymmetric about the equator
\citep{brown10}.  As noted above, the BL term generates opposing
poloidal flux at mid-latitudes near the surface, producing multi-polar
structure and enhancing dissipation.  This plays an essential role in
establishing the cycles of Fig \ref{fig:bflys}$d$-$e$ and is evident
in movies of the mean-field evolution.  The typical amplitude of the 
mean poloidal field near the surface for $\alpha_0 = 100$ m s$^{-1}$, 
1-2 kG, is much larger than for $\alpha_0 = 0$ ($\sim$ 200 G in 
Case D3) but the overall (3D) magnetic energy is about a factor 
of two smaller.

The transition from steady to cyclic dynamos occurs when the BL term
$S$ competes with the fluctuating emf 
$\left<\vv^\prime \cross \BB^\prime\right>$, where $\vv^\prime = \vv - \left<\vv\right>$ 
and $\BB^\prime = \BB - \left<\BB\right>$.  This in turn occurs when
$\alpha_0$ becomes comparable to the velocity of the convective
motions, $V_c$.  The relevant scale for $V_c$ is the rms value
of the meridional components of $\vv^\prime$, which is about
120 m s$^{-1}$ near the top of the convection zone in Case D3, 
decreasing with depth.

A legitimate question is whether the cyclic dynamo in Fig.\ \ref{fig:bflys}$d$,$e$
is operating in an essentially mean-field, axisymmteric mode.   In other words,
if the BL term were to dominate the generation of mean poloidal field and the 
differential rotation were to dominate the generation of mean toroidal field via
the $\Omega$-effect, then one might expect this 3D simulation to behave very
similarly to an analogous, axisymmetric mean-field model. Under this mean-field
scenario, coupling between the poloidal and toroidal source regions might occur 
through the nonlocality of the BL term, the mean meridional flow, and the
turbulent diffusion $\eta$.  The primary role of the resolved convective 
motions would then be to maintain the mean flows.

In order to assess whether this mean-field scenario 
is indeed a valid interpretation of the cyclic activity shown in Fig.\
\ref{fig:bflys}$d$,$e$, we have initiated another simulation in which
we have artificially suppressed the fluctuating emf.  More specifically,
we have replaced the $\vv \cross \BB$ term in equation (\ref{eq:indy})
with only mean-field induction $\left<\vv\right> \cross
\left<\BB\right>$.  This simulation was restarted from that shown
in Fig.\ \ref{fig:bflys}$d$,$e$ at $t = 8.25$ yrs with the same 
parameters.  The only difference is the absence of the fluctuating emf.

\begin{figure}
\centerline{\epsfig{file=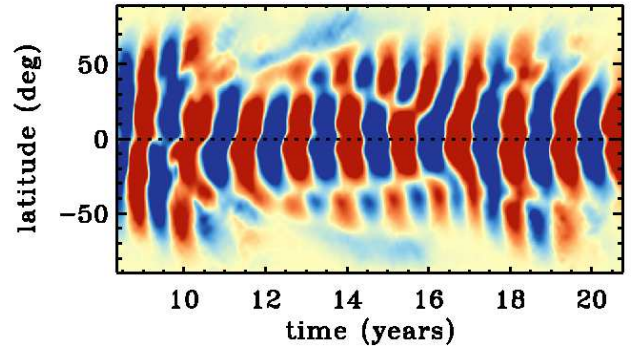,width=3.5in}}
\caption{Mean toroidal field versus latitude and time as in
Fig.\ \ref{fig:bflys} (same radial level, color table, saturation
$\pm$ 4kG) for a simulation with $\alpha_0 = $ 100 m s$^{-1}$, along with 
an artificial suppression of the fluctuating emf 
$\left<\vv^\prime \cross \BB^\prime\right>$ (see text).  
Compare with Fig.\ \ref{fig:bflys}$d$,$e$.
\label{fig:bfly_mf}}
\end{figure}

The dynamo mode changes dramatically, as demonstrated in Fig.\ 
\ref{fig:bfly_mf}.  The non-axisymmetric magnetic field is quickly
dissipated by ohmic diffusion, with the corresponding energy decreasing 
by six orders of magnitude within two years (shear promotes more rapid
dissipation than the nominal 3.6 year diffusion time scale).  The total
magnetic energy is about a factor of three larger than in the progenitor
simulation of Fig.\ \ref{fig:bflys}$d$,$e$ and is dominated by a strong, 
axisymmtric toroidal field that is symmetric about the equator (positive 
parity) and reverses cyclically with a period of about 1.3 yrs.  
The butterfly diagram in Fig.\ \ref{fig:bfly_mf} suggests poleward
propagation but closer inspection of the mean field evolution 
reveals a dynamo wave that propagates toward the rotation axis,
with a cylindrical orientation for the wave front.  This is consistent
with the cylindrical nature of the differential rotation profile
(Fig.\ \ref{fig:D3}$b$) but is strikingly different from the
progenitor simulation in Fig.\ \ref{fig:bflys}$d$,$e$ which
exhibits virtually no mean toroidal field at the equator even
when the parity is positive.  In short, cycles are longer, stronger,
and less solar-like without a turbulent emf.

Thus, the cyclic dynamo in Fig.\ \ref{fig:bflys}$d$,$e$ is not
operating as a mean-field dynamo, or at least not in a naive
sense.  Convection contributes mean field generation and transport
that plays an essential role in shaping the dynamo even for 
$\alpha_0$ as high as 100 m s$^{-1}$.  Parity selection in 
cyclic dynamos is a subtle issue and space limitations 
preclude a thorough discussion here. 
We only remark that mean-field BL models have demonstrated
that the relatively homogeneous nature of turbulent pumping
\citep{guerr08}, the antisymmetric structure of a supplementary 
$\alpha$-effect operating near the base of the convection zone
\citep{dikpa01c}, and efficient turbulent mixing that peaks in the upper
convection zone \citep{hotta10} can all help promote negative (dipolar) 
parity.  These mean-field results are loosely consistent with the 3D convection
simulations reported here but warrant a more careful investigation which we 
reserve for future work.

In summary, we have shown that flux emergence can promote cyclic magnetic
activity in a convective dynamo by means of the Babcock-Leighton mechanism.
Although this result is to some extent anticipated by mean-field dynamo
models, we have demonstrated it for the first time in a global convection 
simulation.  The BL mechanism is not required to achieve cycles in convective
dynamo simulations but, as we have shown here, it may help shape cycle properties 
such as the period, amplitude, and equatorward propagation of toroidal flux.

However, achieving cyclic activity through the BL mechanism is not
easy.  To make a significant impact on the dynamo, the amplitude of
the BL $\alpha$-effect must be comparable to the convective
velocity, $V_c$.  In particular, we find that $\alpha_0 > $ 10 m s$^{-1}$
is required to induce cyclic activity when $V_c \sim$ 100 m s$^{-1}$ in the
upper convection zone.  By comparison, typical values of $\alpha$ used
in mean-field BL dynamo models of the solar cycle are less than 
1 m s$^{-1}$ \citep[e.g.][]{dikpa09,charb10}.

The relatively large value of $\alpha_0$ used here, together with the
strong shear $\vert \del \Omega\vert$ and the efficiency of convective
transport, can likely account for the very short cycle period in this
young, rapidly-rotating Sun ($\Omega = 3 \Omega_\odot$).  Might other
parameters produce a 22-yr period comparable to the solar cycle?  This
remains to be seen.  Whether the cycle period is regulated by a dynamo
wave or flux transport, the large value of $\alpha_0$ needed to
promote cyclic activity ($> 10$ m s$^{-1}$) and the short convective
turnover time scale ($\sim $ 20 days) may favor relatively short cycle
periods.  Longer cycle periods can generally be achieved in mean-field
BL dynamo models by reducing the efficiency of poloidal flux transport
\cite[e.g][]{dikpa09,charb10} but we do not have that freedom here.  Here the
efficiency of transport is set by the convection, which in turn is an
output of the simulation, regulated mainly by factors such as the stellar
mass and luminosity that are set by observations.  This implies either
that the mean field generation and transport by solar convection is
much less efficient than in the models considered here (due possibly
to dynamical quenching of the turbulent $\alpha$ and $\beta$-effects
or overestimation of the convective velocity), or that the solar
dynamo does not adhere to a simple BL paradigm.

\acknowledgements
We thank Mausumi Dikpati and Matthias Rempel for insightful discussions
and comments on the manuscript and we thank the anonymous referee for a 
prompt and constructive report.  This work is supported by NASA grants 
NNH09AK14I (SR\&T) and NNX08AI57G (HTP) and computing resources were 
provided by the NASA High-End Computing (HEC) Program and the NSF-sponsored 
Teragrid resources at NICS and TACC.  B.P.\ Brown is supported in part by 
NSF Astronomy and Astrophysics postdoctoral fellowship AST 09-02004.  
NCAR is sponsored by NSF and CMSO is is supported by NSF grant PHY 08-21899.

\end{document}